\documentclass[twocolumn,preprintnumbers,amsmath,amssymb]{revtex4-1}
\usepackage{epsfig}
\usepackage{color}
\usepackage{amsmath}
\usepackage{multirow}

\begin{document}

\title{Seeding Approach to nucleation in the NVT ensemble: the case of bubble cavitation in overstretched Lennard Jones fluids}

\author{P. Rosales-Pelaez$^1$, I. Sanchez-Burgos$^1$, C. Valeriani$^2$, C. Vega$^1$ and E. Sanz$^1$}
\affiliation{$^1$Departamento de Qu\'{\i}mica F\'{\i}sica,
Facultad de Ciencias Qu\'{\i}micas, Universidad Complutense de Madrid,
28040 Madrid, Spain\\
$^2$ Departamento de Estructura de la Materia, F\'{\i}sica Termica y Electronica,
Facultad de Ciencias  F\'{\i}sicas, Universidad Complutense de Madrid,
28040 Madrid, Spain}

\begin{abstract}

Simulations are widely used to study nucleation in first order phase transitions due to the fact
that they have access to the relevant length and time scales. 
However, simulations face the problem that nucleation is an activated process. 
Therefore, rare event simulation techniques are needed to promote the formation of the critical nucleus. 
The Seeding method, where the simulations are started with the nucleus already formed,
has proven quite useful in efficiently providing estimates of the nucleation rate for a wide range of orders of magnitude. 
So far, Seeding has been employed in the NPT ensemble, where the nucleus either grows
or redissolves. Thus, several trajectories have to be run in order to find the thermodynamic conditions that
make the seeded nucleus critical. Moreover, the nucleus lifetime is short and the statistics for obtaining its
properties is consequently poor. To deal with these shortcomings we extend the Seeding method to the NVT ensemble. 
We focus on the problem of bubble nucleation in a mestastable Lennard Jones fluid. We show that, in the NVT ensemble, 
it is possible to equilibrate
and stabilise critical bubbles for a long time. 
The nucleation rate inferred from NVT-Seeding is fully consistent with that coming from NPT-Seeding. 
The former is quite suitable to obtain the nucleation rate along isotherms, whereas the latter 
is preferable if the dependence of the rate with temperature at constant pressure is required. 
Care should be taken with finite size effects when using NVT-Seeding. 
Further work is required to extend NVT seeding to other sorts of phase transitions.

\end{abstract}

\maketitle

\section{Introduction}
The study of the onset of first order phase transitions by means of computer simulations 
has received a great deal of attention \cite{JCP_1992_96_04655,searJPCM2007,anwarAnge2011,reviewMichaelides2016,coluzza2017perspectives}. 
Simulations is a very suited tool to study 
this phenomenon given that the nucleation of the stable phase in the parent mestastable phase 
typically 
entails hundreds of molecules and takes a few nanoseconds \cite{kelton}. These are time and length scales accessible 
to simulations but difficult to probe in experiments.

One of the main difficulties that simulations face in nucleation studies is the activated nature of such process. 
As a consequence, there is a huge timescale difference between the duration of the nucleation process itself and 
the time required for a nucleation event to start. To deal with this problem, different rare event simulation
techniques have been used to promote the formation of the  nucleus \cite{JCP_1992_96_04655,searJPCM2007,anwarAnge2011,reviewMichaelides2016,PRL_2005_94_018104,haji-akbariPNAS2015,bolhuis2002transition,PRL_94_2005_235703,quigley:154518,PNAS_2002_99_12562}.

A rather recent approach, named Seeding, consists in directly starting the simulation from a configuration 
where the nucleus of the stable phase is already formed \cite{baiJCP2006,carignano,knottJACS2012,jacs2013,seedingvienes}. 
This approach is not fully rigorous as it relies on the validity of Classical Nucleation 
Theory \cite{kelton,ZPC_1926_119_277_nolotengo,becker-doring} and on a judicious choice for the criterion used to determine 
the nucleus size \cite{seedingbaronNaCl}. 
Despite the lack of rigour, this method has proven successful in predicting nucleation rates in
crystal nucleation of hard spheres, Lennard Jones spheres, water or sodium chloride \cite{seedingvienes}. 
More recently, we have shown that Seeding is also successful to study vapor cavitation \cite{seedingNpT}. 

In all the works above mentioned, Seeding has been applied at constant pressure and temperature. In this ensemble, 
the inserted nucleus either grows or redissolves depending on whether its size is post or pre-critical at the
simulated pressure and temperature. This implies that in order to find the conditions at which the inserted 
nucleus is critical one has to 
run several trajectories.
Moreover, the lifetime of the nucleus is quite short because it either grows or shrinks quite quickly. 
Therefore, the statistics when computing its properties is quite poor. 

To deal with
these shortcomings of Seeding in the NPT ensemble  (NPT-Seeding) in this work we develop a Seeding variant at constant volume, temperature and number of molecules: NVT-Seeding. 
Inspired by previous work where nuclei are generated and stabilised at constant volume in the grand canonical 
ensemble \cite{macdowell2006nucleation,koss2017free,statt2015finite,finitesizenucreguera}, 
we consider the possibility of stabilising a seeded nucleus in the NVT ensemble. 
To do this we choose to investigate the liquid-to-vapor transition, 
which is of great interest both in nature and industry
\cite{shusser1999explosive,shusser2000kinetic,toramaru1989vesiculation,massol2005effect,brennen2014cavitation,suslick1990sonochemistry,suslick1997chemistry,kahler2015cavitation}. 
We have recently applied NPT-Seeding to study bubble nucleation in a system composed of Lennard Jones particles \cite{seedingNpT} 
and we found a good accordance with other studies where rigorous 
rare event techniques were used \cite{wang2008homogeneous,meadley2012thermodynamics}.

In this work we show that NVT and NPT-Seeding give the same 
bubble nucleation rate. In NVT-Seeding only one trajectory is needed to obtain the rate because critical bubbles are 
spontaneously equilibrated from a generated cavity. Moreover, the critical bubble remains stable for the simulation duration, 
which makes it possible to accurately obtain its properties.   
The system size imposes some applicability limitations, though. 
On the one hand, finite size effects may appear when generating large cavities if the 
remaining liquid is not enough to define bulk thermodynamic properties in it. On the other hand, too small cavities redissolve and cannot be equilibrated. 
We find that NVT-Seeding is quite suitable to obtain the nucleation rate along an isotherm. 
Further work is required to prove if NVT-Seeding is also valid to study other sorts of phase transitions such as crystallization or condensation.

\section{Seeding}
The Seeding method consists in using the expressions given by Classical Nucleation Theory (CNT) \cite{kelton,ZPC_1926_119_277_nolotengo,becker-doring}
to estimate with computer simulations parameters that characterize nucleation, such as nucleation free energy barriers, 
interfacial free energies and, most importantly, nucleation rates. In our particular case, we study 
the nucleation of vapor bubbles in an overstretched fluid. The CNT rate expression we employ is: 
\begin{equation}
\label{rateeq}
J= \rho_l \sqrt{\frac{\Delta P R_{c}}{\pi m}} \exp \left (-\frac{2 \pi R_{c}^3 \Delta P}{3 k_B T} \right ), 
\end{equation}
where $R_c$ is the radius of the critical bubble (which is assumed to be spherical), $\rho_l$ is the density
of the parent liquid phase, $\Delta P$ is the pressure difference between the liquid and the bubble and $k_B$ is the 
Boltzmann constant. 
The exponential 
pre-factor is an expression provided by Blander and Kats \cite{Blanderino}. 
In a recent publication, we showed that such expression gives the same result as 
two other alternative ways of computing the kinetic pre-factor \cite{seedingNpT}. We use here 
the Blander and Kats expression because it is the handiest one \cite{seedingNpT}.
In summary, to obtain the nucleation rate, one must just obtain via computer simulations $R_c$, $\Delta P$ and $\rho_l$. 

In a previous publication \cite{seedingNpT} we showed how to obtain these parameters with simulations of a bubble surrounded by 
liquid at constant pressure and temperature (NPT ensemble). 
Here, we obtain $\Delta P$ and $R_c$ at constant volume and temperature (NVT ensemble).
A critical difference between both ensembles is the bubble lifetime. 
At constant pressure a bubble is either sub-critical or post-critical and it will accordingly
either dissolve or grow. As a consequence, the bubble lifetime is short. 
At constant volume, however, we expect that longer lifetimes are accessible given that a liquid with a bubble under periodic boundary
conditions represents a minimum in the (Helmholtz) free energy landscape \cite{macdowell2006nucleation,finitesizenucreguera}.
Long bubble lifetimes improve the statistics in the calculation of $\Delta P$ and $R_c$ in the simulations. 
Moreover, as discussed below, the preparation of the seeding configuration (a bubble surrounded by liquid) is easier 
in the NVT ensemble. 

\section{Simulation details}
To validate our new Seeding approach to study cavitation in the NVT ensemble, we carry out
computer simulations of the truncated and force-shifted Lennard-Jones
(TSF-LJ) potential \cite{wang2008homogeneous}, a model for which
bubble cavitation has been previously studied \cite{wang2008homogeneous,tanaka2015simple,meadley2012thermodynamics,seedingNpT}:
\begin{equation}
U_{TSF-LJ}(r) = U_{LJ}(r) - U_{LJ}(r_{c}) - (r-r_{c})U'_{LJ}(r_{c}), 
\end{equation}
where $U_{LJ}(r)$ is the 12-6 Lennard-Jones potential and $U'_{LJ}(r)$ its
first derivative. The interaction potential is truncated and shifted at  $r_{c}=2.5 \sigma$,
being $\sigma$ the particle's diameter  and $\epsilon$ the depth of the un-truncated Lennard-Jones potential.
In what follows, we will use reduced units, expressing all physical variables
in terms of $\sigma$, $\epsilon$ and $m$, where $m$ is the mass of the
particles. Reduced variables are indicated with an asterisk: 
$T^{*}=k_{B}T \epsilon^{-1}$, $P^{*}=P \cdot \sigma^{3}
\epsilon^{-1}$, $\rho^{*}=\rho \cdot \sigma^{3}$, $t^{*}=t \left ( \epsilon
m^{-1} \sigma^{-2} \right )^{1/2}=t / \tau$, $\gamma^{*}=\gamma \cdot \sigma^2 \epsilon^{-1}$ and $J^{*}=J \sigma^3 \tau$.
All simulations are performed at $T^* = 0.785$, which is the coexistence temperature at $P^*=0.026$. 

All simulations have been performed using the Molecular Dynamics (MD) LAMMPS package \cite{lammps_program},
applying cubic periodic boundary conditions and
integrating the equations of motion with a leap-frog algorithm \cite{leapfrog}
with a time-step of $\Delta t^* = 0.0012$.

To carry out NVT simulations, the temperature was fixed using the Nos\'e-Hover thermostat \cite{JCP_1984_81_00511};
and when NPT simulations were run, both the temperature and pressure were held constant via a
Nose-Hover thermostat and barostat \cite{JCP_1984_81_00511}
with relaxation times $\tau_{T}=0.46 \tau$ and $\tau_{P}=4.6 \tau$ respectively.

\section{Results}

The Seeding method in the NVT ensemble consists in generating an initial configuration with a bubble 
surrounded by liquid from which a long NVT simulation is run to compute $R_c$
and $\Delta P$. Then, Eq. \ref{rateeq} is used to estimate the nucleation rate.
In the following we give details of the NVT-Seeding calculations and present and discuss the results. 

\subsection{Phase diagram}

\begin{figure}[h]
\begin{center}
	\includegraphics[width=0.85\linewidth]{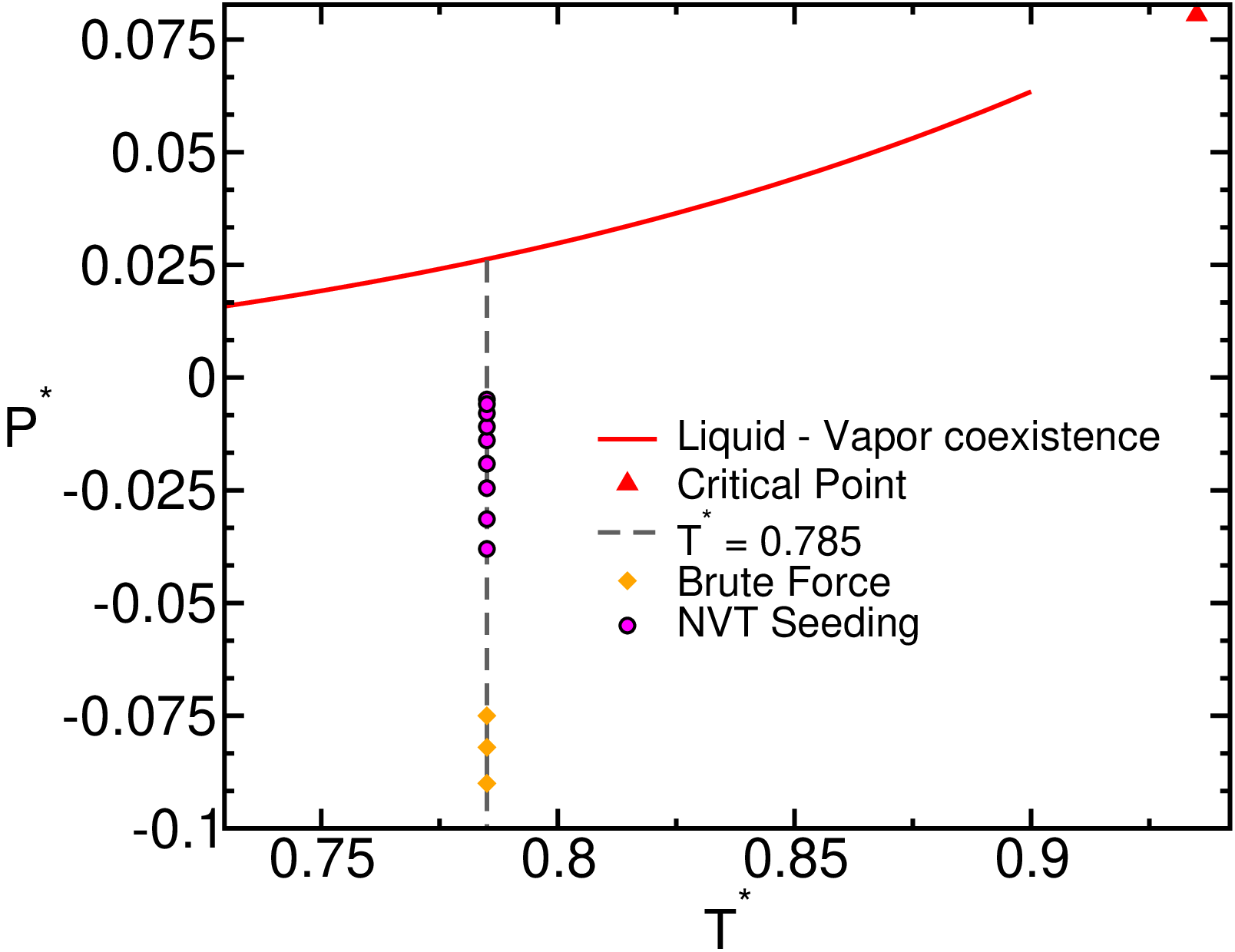}
	\caption{Temperature-pressure phase diagram of the Lennard Jones model under study. In red, the coexistence line (the triangle corresponds to the critical point). The dashed
vertical line indicates the isotherm under investigation. Pink circles correspond to the points where NVT-Seeding simulations were carried out and orange diamonds to those where the
bubble cavitation rate was determined by brute force molecular dynamics.}
\label{phasediag}
\end{center}
\end{figure}

In Fig. \ref{phasediag} we show the temperature-pressure plane of the equilibrium phase diagram 
of the employed Lennard-Jones model \cite{wang2008homogeneous}. The red curve is the coexistence line and 
the triangle indicates the location of the critical point. 
The dashed line corresponds to the studied isotherm. Pink circles indicate the state points where
we did NVT-Seeding simulations whereas the diamonds correspond to brute force molecular dynamics calculations 
of the nucleation rate. 

\subsection{Bubble radius for an instantaneous configuration}
\begin{figure}
\begin{center}
	\includegraphics[width=0.85\linewidth]{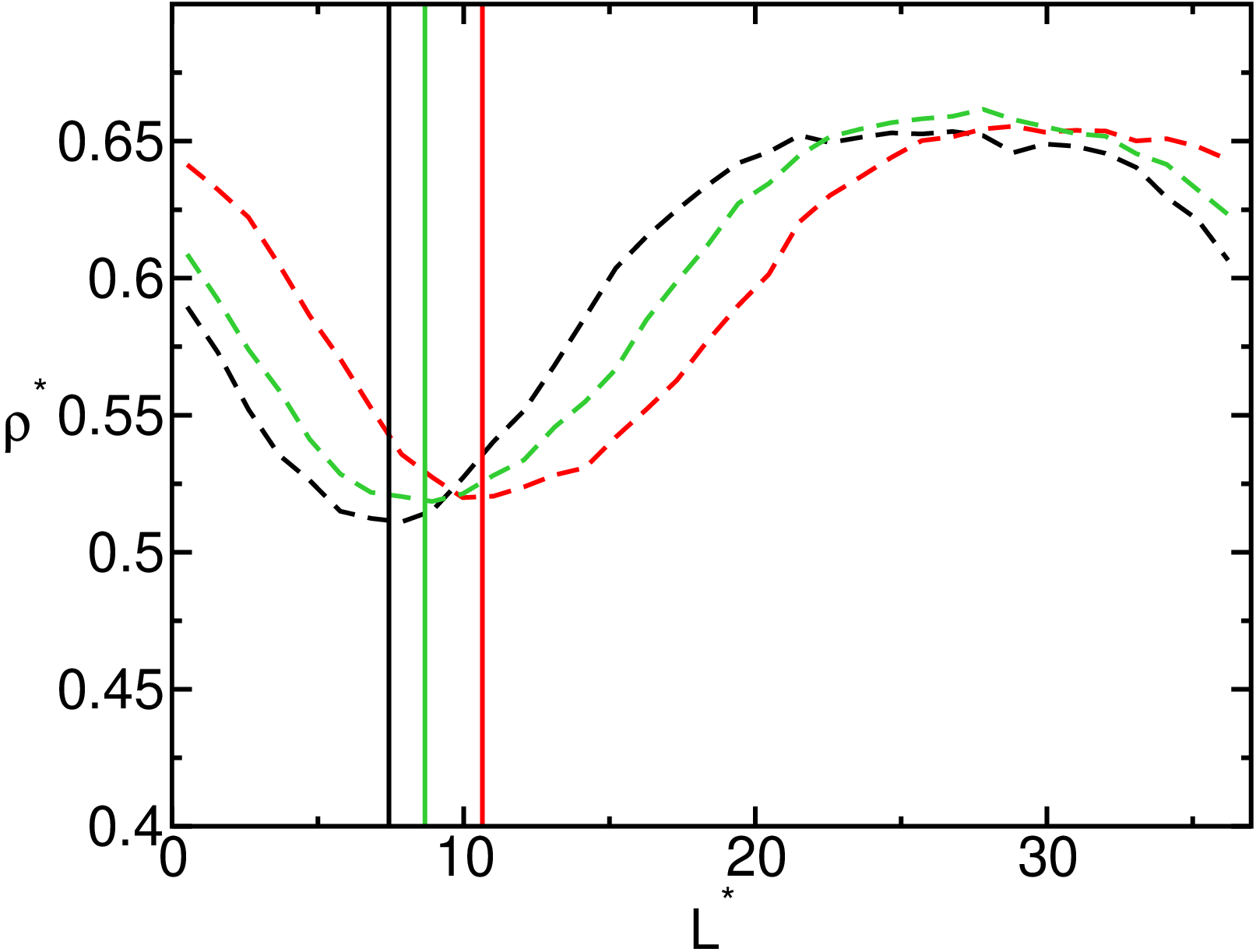}(a)
	\includegraphics[width=0.85\linewidth]{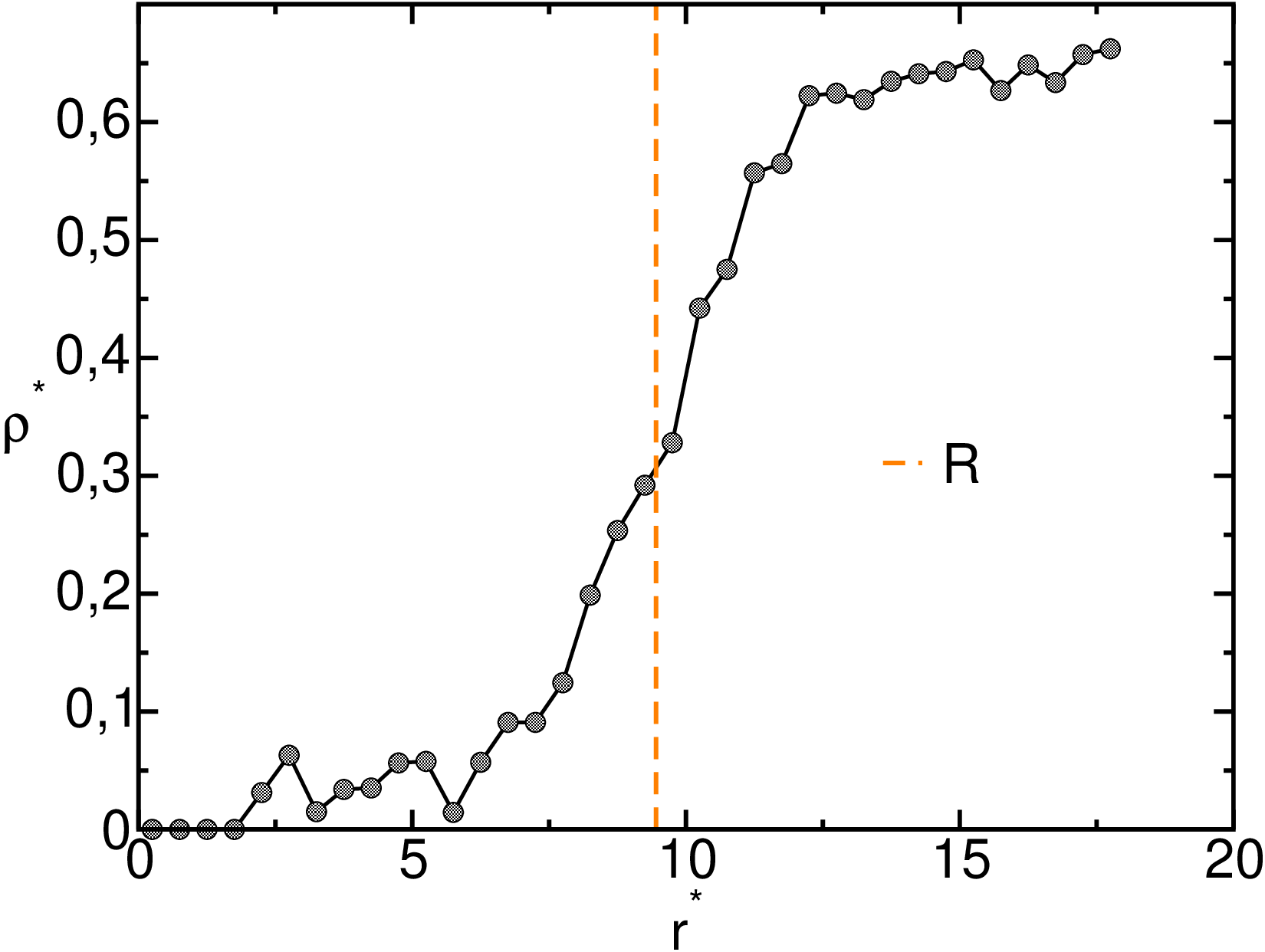}(b)
	\caption{(a) Density profiles along the $x$ (black), $y$ (red) and $z$ (green) 
	coordinates of the simulation box. The bubble centre along each coordinate
	is located at the corresponding density profile minimum (indicated with vertical lines in the figure). (b) Radial density profile centred at
	the position identified in (a). The vertical dashed line indicates the bubble radius obtained, as described in our previous work \cite{seedingNpT},
	as the radius for which the density is average between that inside the bubble and that of the surrounding liquid.}
\label{bubblecenter}
\end{center}
\end{figure}

To obtain the bubble radius we compute spherically symmetric density profiles from the bubble center.
A substantial difference with respect to our previous seeding work in the NPT ensemble is that, when simulated at constant volume,
the bubble can drift away during the 
course of the simulation (at constant pressure the bubble lifetime is not long enough 
for drifting). Therefore, we must first find the bubble center for each configuration. 
We identify the bubble centre coordinates with the density profile minima 
along each cartesian coordinate (see Fig. \ref{bubblecenter} (a)).  
From the obtained center coordinates, a radial density profile is computed. 
As in our previous Seeding work \cite{seedingNpT}, 
the bubble radius, $R$, is identified with the position at which the density is 
average between that of the liquid and that in the interior of the bubble ($R^{ED}$ in Ref. \cite{seedingNpT}). 
See Fig. \ref{bubblecenter}(b) for an example of a density profile. 
We do not repeat here the details on the calculation of the bubble radius from density profiles, which are profusely described in Ref. \cite{seedingNpT}.

\subsection{Preparation of the initial configuration}
\label{preparation}

To induce a bubble of certain radius we take an equilibrated configuration of
the liquid of density $\rho_l$ and $N_0$ particles at the temperature of interest
and we randomly subtract $N_{-}$ particles. We estimate the number of particles needed to be 
erased in order to obtain a bubble with a certain intended 
radius, $R_{i}$, as:
\begin{equation}
\label{intended}
N_{-} = \frac{4 \pi \rho_{l}}{3} R_{i}^{3}.
\end{equation}
Then, we are left with $N_T$ particles 
and we switch on a spherically symmetric step-like repulsive potential, $U_{rp}$, at a given point of the simulation box 
to create a cavity in the fluid:
\begin{equation}
\label{Pozo_EQ}
	U_{rp} =  \frac{\epsilon_{rp}}{2} \sum_{i=1}^{i=N_T} \left [ 1 - \tanh \left ( \frac{r_{i-rp}-r_{rp}}{\alpha_{rp}} \right )\right ].
\end{equation}
Here, $r_{rp}$ and $\epsilon_{rp}$ are the radius and the height of the repulsive potential respectively, 
$r_{i-rp}$ is the distance between the centers of the repulsive potential and particle $i$ and $\alpha_{rp}$ controls the steepness
of the repulsive step (we use $\alpha_{rp}^*=0.005$). 

The precise value of the parameters of the repulsive potential are unimportant as long as the cavity is formed. 
For instance, we carried out simulations with a repulsive potential of radius 
$r_{rp}=5.0 \sigma$ and $r_{rp}=3.5 \sigma$, both with $\epsilon = 1.0 k_BT$ and
the resulting bubbles were identical.
\textcolor{black}{
This is due to the fact that the system minimizes the Helmholtz free energy, F, at constant N, V and T, which guarantees that the bubble size converges to an equilibrium value regardless the initial size of the cavity generated.}
$\epsilon_{rp}$ must be high enough 
to repel liquid particles from the repulsive potential area, but 
not too high to exert too strong forces in the particles leaving the repulsive area (otherwise
pressure waves appear and it takes longer to equlibrate the system). 
We found $\epsilon_{rp} = 1.0 k_BT$ to be a good value.

\begin{figure}[h]
\begin{center}
	\includegraphics[width=0.85\linewidth]{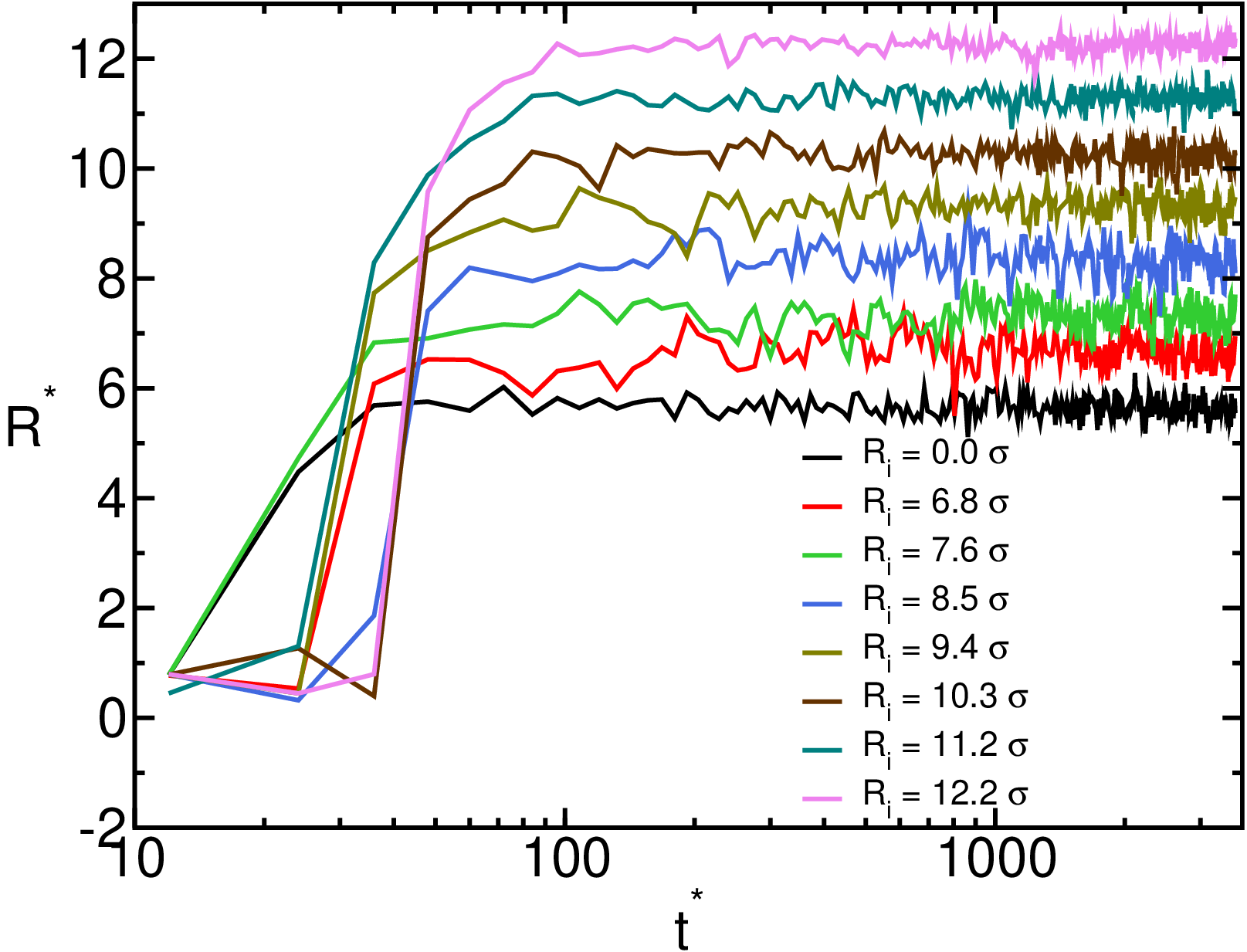}(a)
	\includegraphics[width=0.85\linewidth]{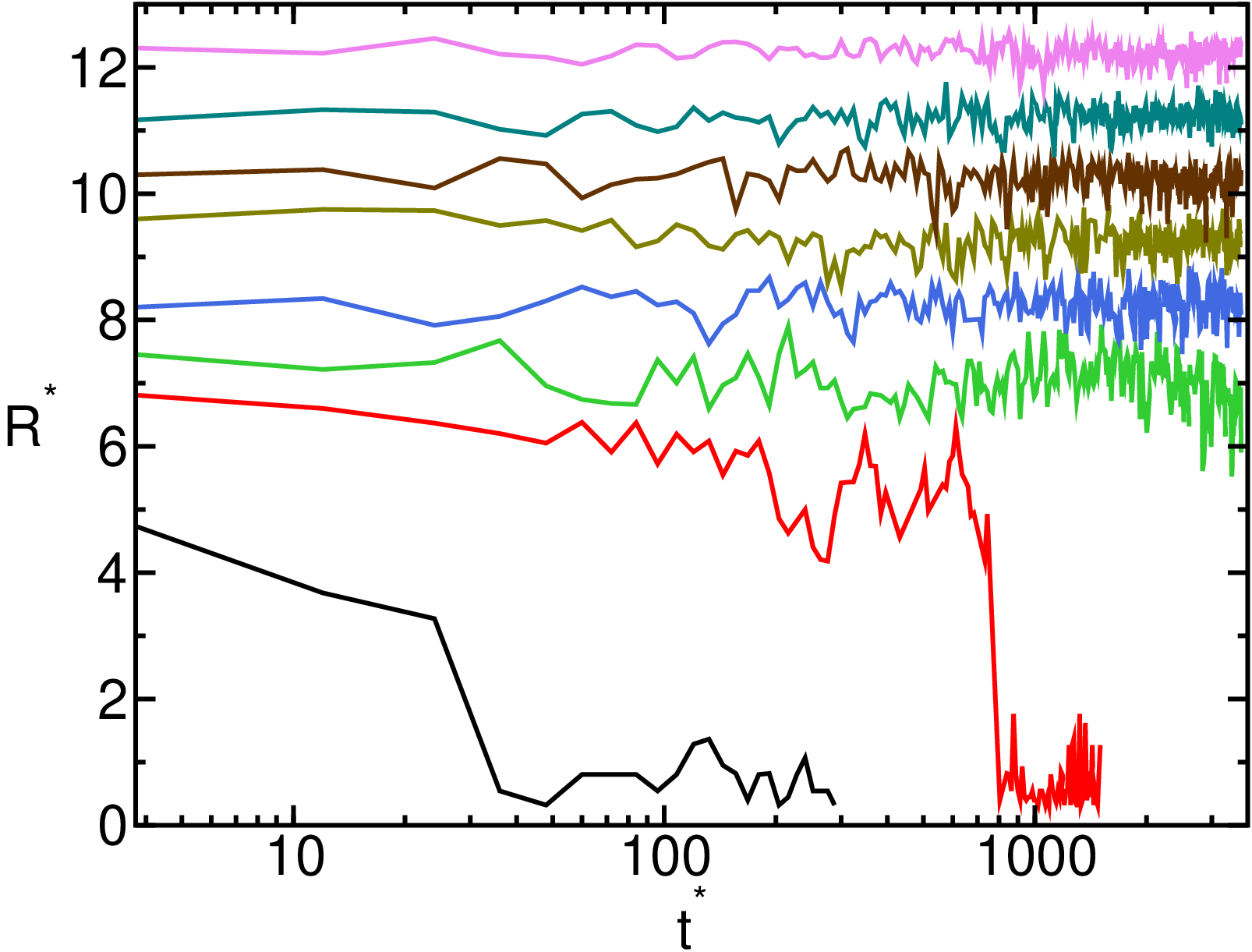}(b)
	\caption{(a)Radius of the cavity as a function of time. A repulsive step-like potential is switched on at the beginning of the simulation. 
	Different cavity radii are obtained by subtracting different amount of particles to aim at a certain 
	intended cavity radius ($R_i$ in the legend) according to Eq. \ref{intended}. The simulation box edge in all cases is $L = 36.7307 \sigma$. 
	(b) Continuation of the runs shown in (a) but with the repulsive potential switched off.}
\label{Hueco_GR}
\end{center}
\end{figure}

In figure \ref{Hueco_GR}(a) we show the time evolution of the radius of the induced cavity. The repulsive potential
is switched on at time 0. All these simulations are performed
with a repulsive potential of $r_{rp}=5 \sigma$ and $\epsilon_{rp} = 1.0 k_BT$.
Cavities with different radii have been induced by erasing a different amount of particles from the simulation box. 
The corresponding intended radius, $R_i$ in Eq. \ref{intended}, is reported in the figure legend. 
The correspondence between the radius of the generated cavity and $R_i$ is quite good. Obviously, when no particles
are removed ($R_i^*=0$) the cavity radius is close to $r_{rp}$. 

From the previous step we have configurations with the same simulation box edge, $L$, different number
of particles and a cavity with a certain radius. In table  \ref{tabla} we give details on the 
systems employed in this work. Those with $L^*=36.73$ correspond to the simulations 
shown in Fig. \ref{Hueco_GR}.

\subsection{Critical bubble radius}
\label{cbr_sec}

We now switch off the repulsive potential and let the system equilibrate. In Fig. \ref{Hueco_GR}(b) we show 
the continuation of the simulations shown in Fig. \ref{Hueco_GR}(a) but switching off the repulsive potential at time 0. 
In most cases the cavity generated due to the repulsive potential remains stable throughout the simulation, i. e., we got equilibrated
bubbles
\textcolor{black}{whose size is dictated by a minimisation of F in the 
system.
The F minimum could be a local one and}
a cylindrical bubble pipe could form when the simulation is run for long times \cite{schrader2009simulation,binder2012beyond}, 
but we have not come across this problem. 
For $R_i^*=0$ the cavity immediately redissolved and we did not get any stable bubble. 
For $R_i^*=6.8$ the bubble remained stable for about $t^*=$750 and then redissolved. This is enough time to compute 
the bubble properties, as we show later on.  

\begin{figure}[h]
\begin{center}
	\includegraphics[width=0.85\linewidth]{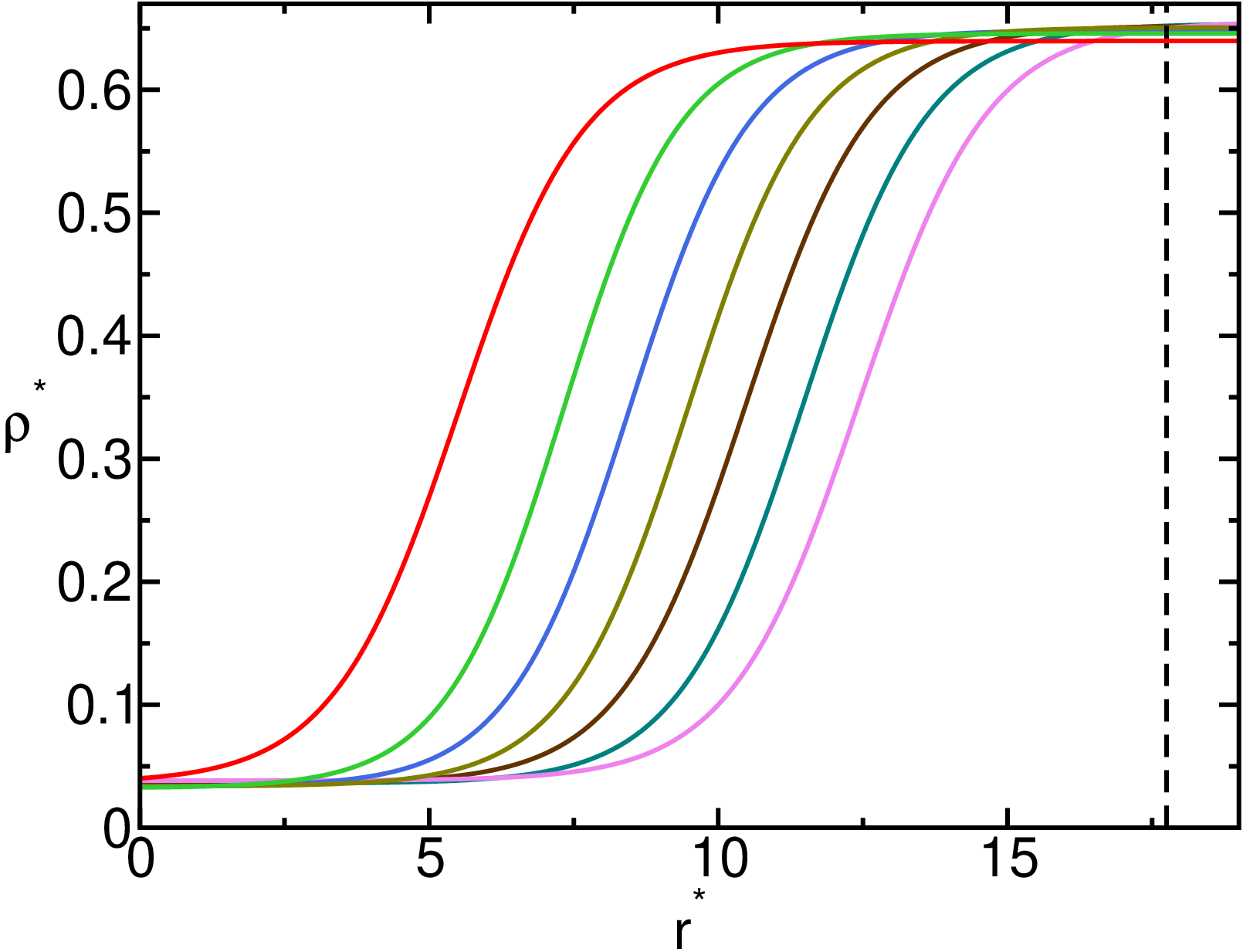}(a)
	\includegraphics[width=0.85\linewidth]{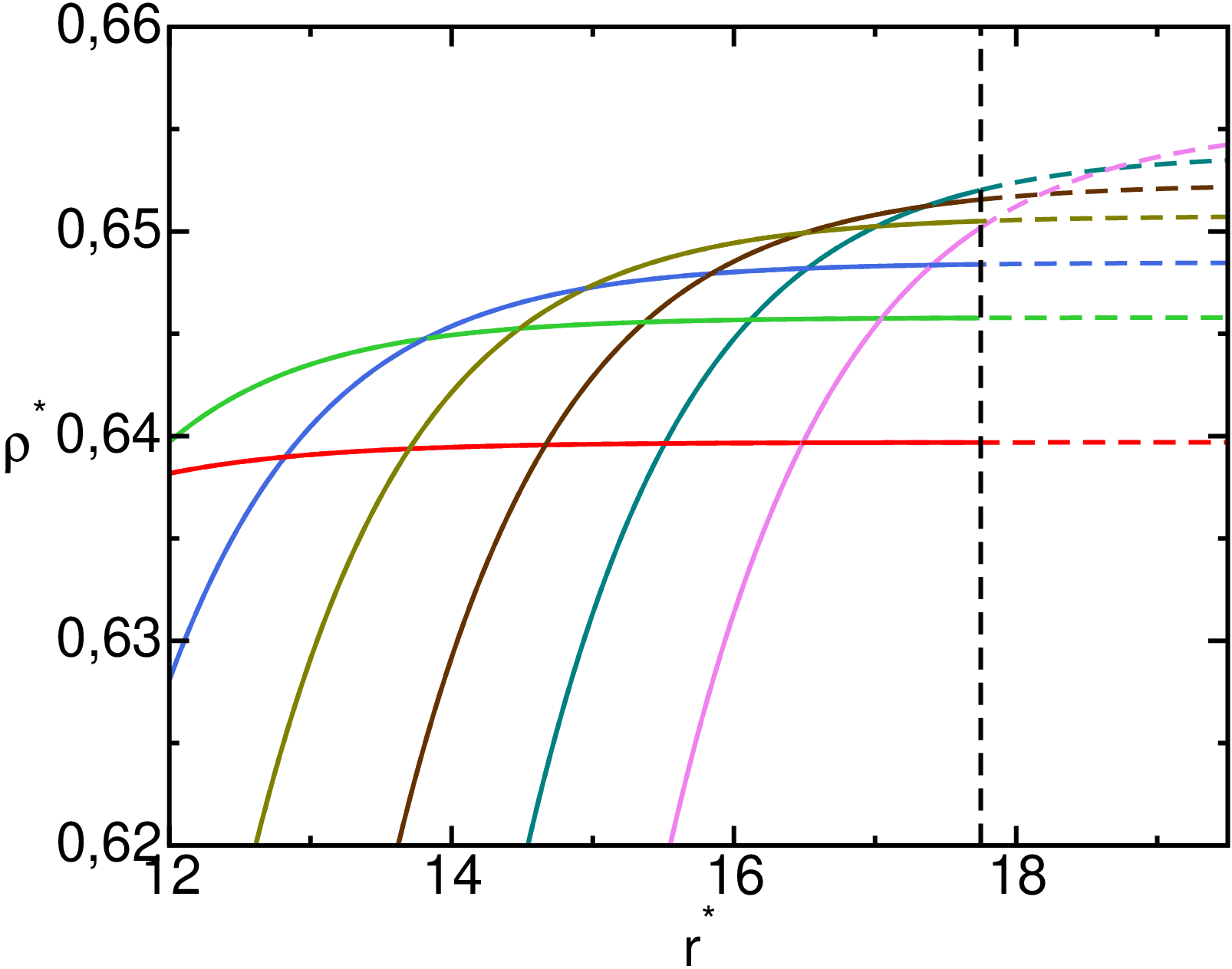}(b)
	\caption{(a) Fits to the radial density profiles of the stabilised bubbles (Eq. \ref{fit}). 
The color code is the same as in Fig. \ref{Hueco_GR}. 
The vertical dashed line indicates a distance of half the edge of the simulation box. Beyond that distance the
	fits are an extrapolation and are shown in dashed curves. In part (b) 
a zoom close to that distance is shown.}
\label{densityprofiles}
\end{center}
\end{figure}

Not all trajectories with a stable bubble are suitable to obtain the bubble properties: it should also 
be possible to define bulk properties in the surrounding liquid. 
If the bubble is too big as compared to $L$, a too thin liquid layer
will separate one bubble from its periodic replica. 
In Fig. \ref{densityprofiles}(a) we show radial density profiles  starting from the bubble centre of the simulations 
with a stable bubble.
The curves shown in the figure are in reality fits of superimposed density profiles to the following sigmoid function:  
\begin{equation}
	\rho(r) = \frac{\rho_{v,dp} + \rho_{l,dp}}{2} + \left ( \frac{\rho_{l,dp} - \rho_{v,dp}}{2} \right ) \cdot \tanh \left [(r - R_{c}) / \alpha \right ],
	\label{fit}
\end{equation}
where $\rho_{v,dp}$ and $\rho_{l,dp}$ are
the densities of the vapor and the liquid phases obtained with the density profile (dp) fit respectively, 
$\alpha$ is a parameter related to the width of the interfacial region and $R_{c}$, our definition of the critical bubble radius, 
is the distance at which the density is the average between 
both phases (i. e.
the ``equi-density'' critical radius in Ref. \cite{seedingNpT}):
\begin{equation}
	\rho(R_c)=\frac{\rho_{v,dp}+\rho_{l,dp}}{2}
\end{equation}
The fit parameters $R_c$, $\rho_{l,dp}$ and $\rho_{v,dp}$ are reported in Table \ref{tabla}.

Due to periodic boundary conditions 
the density profiles can only be computed until $L/2$.  
Therefore, the fits shown in Fig. \ref{densityprofiles} are an extrapolation beyond $L/2$ (shown with dashed curves). 
As it can be seen in part (b) of the figure, 
for the two biggest bubbles, the liquid density has not 
converged by $L/2$, indicated by a vertical dashed line in the figure. 
In these cases the liquid does not reach a uniform bulk density along the directions perpendicular to the box sides.
We can nonetheless define the liquid density via $\rho_{l,dp}$, which is a parameter of the fit 
above. 

\begin{figure}[h]
\begin{center}
	\includegraphics[width=0.85\linewidth]{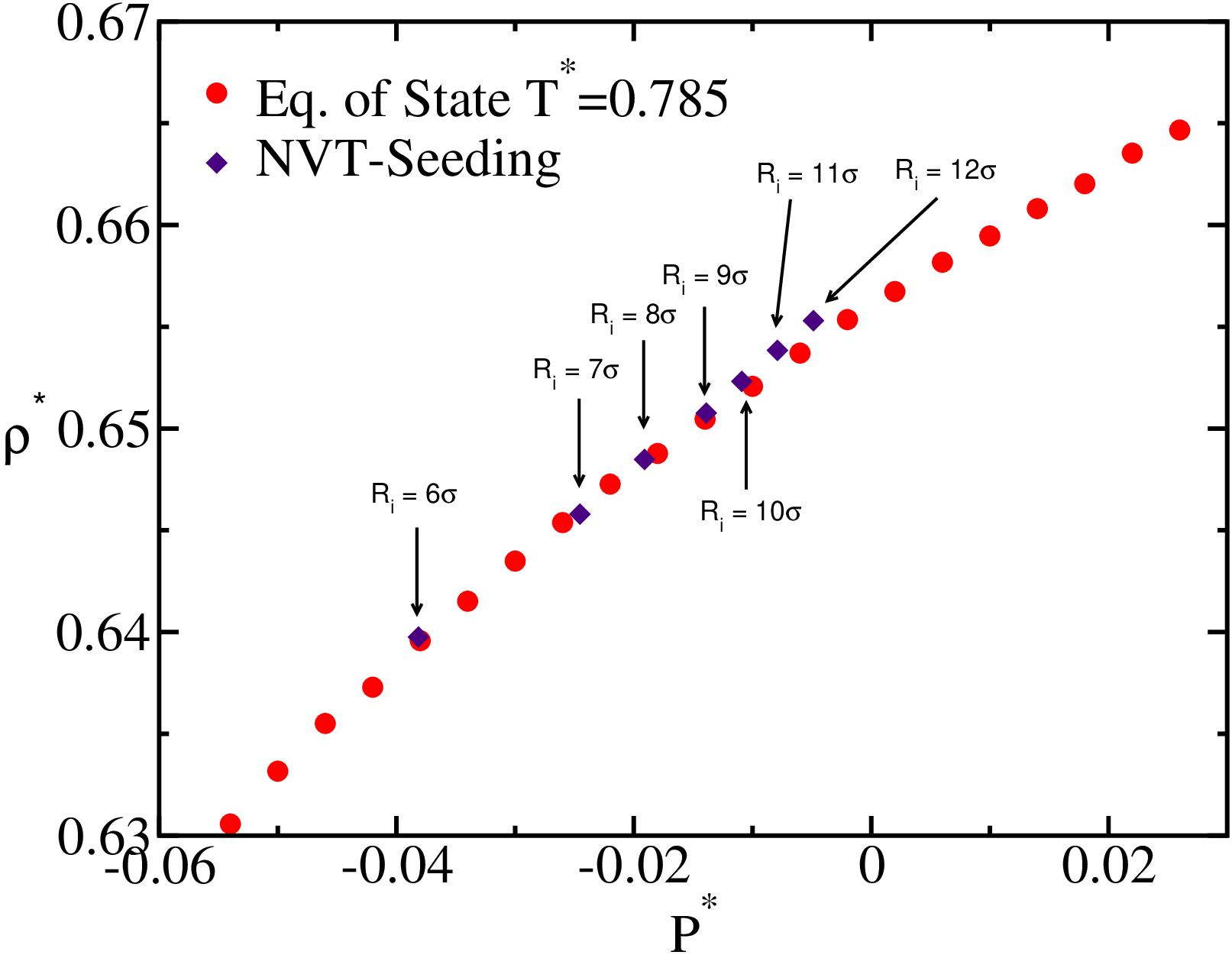}
	\caption{Bulk liquid density equation of state (red dots) compared with the density of the liquid surrounding the bubble, $\rho_{l,dp}$, versus the
	virial pressure of the whole system (purple diamonds).}
\label{vm}
\end{center}
\end{figure}

In Fig. \ref{vm} we show $\rho_{l,dp}$ as a function of the pressure obtained in the simulations
with the virial equation (purple diamonds).
These are compared with the red dots, which have been obtained in bulk liquid simulations. 
The agreement is quite satisfactory,  
\textcolor{black}{which means that the nucleus is small as compared to
the whole system and its
presence does not affect the liquid density-pressure relation in any measurable manner. Thus, the thermodynamic state of the bulk liquid phase that surrounds the
bubble can be estimated from the simulation that contains the bubble. 
} Only small deviations between the bulk equation of state (red 
points) and that 
coming from the bubble simulations (purple diamonds) 
are present in the two biggest bubbles, 
which is perhaps expected from the afore-mentioned finite size effects
in these cases.

\begin{figure}[h]
\begin{center}
        \includegraphics[width=0.85\linewidth]{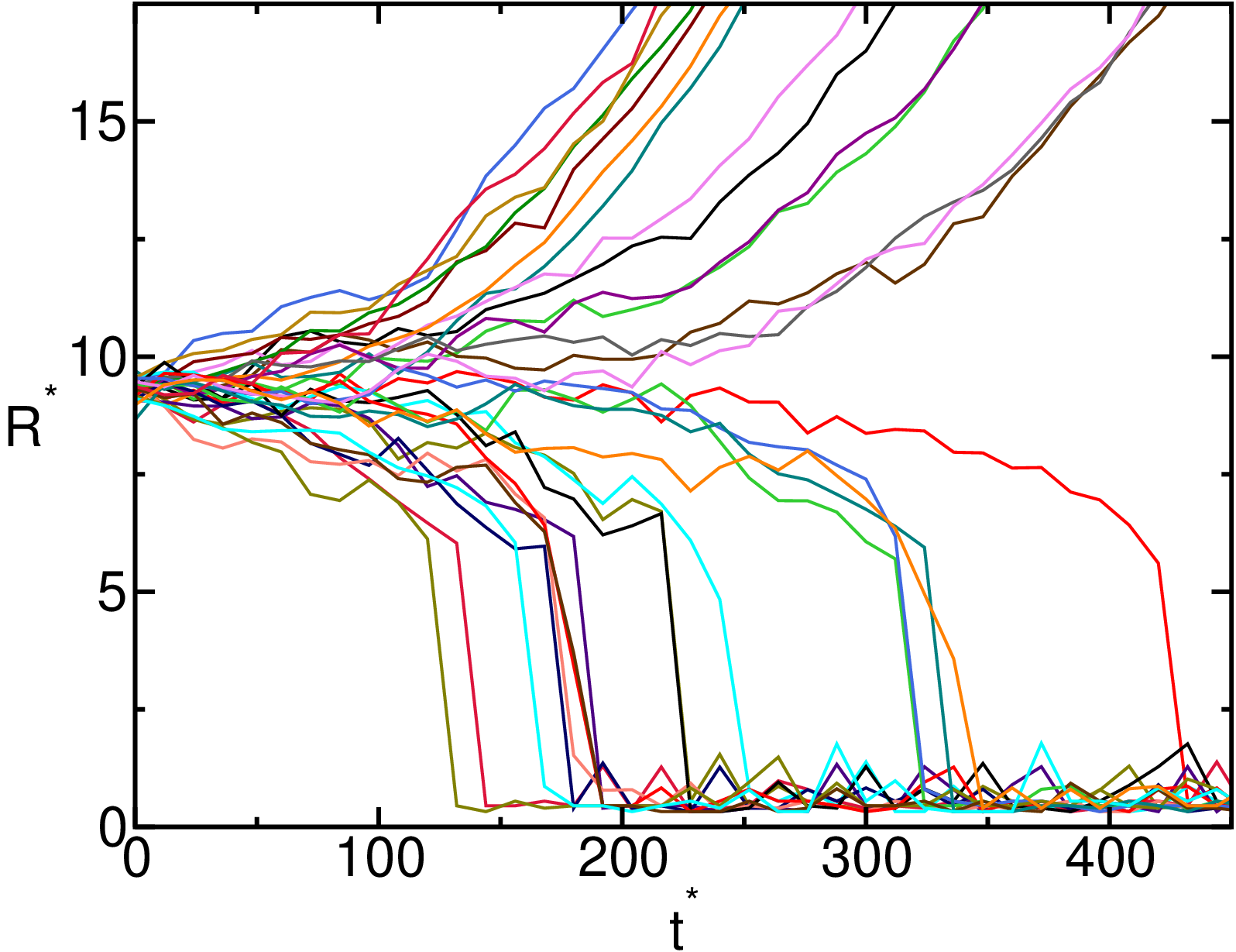}
	\caption{Bubble radius time evolution in the NPT ensemble 
	for 30 configurations taken from an NVT simulation of an equilibrated bubble (that corresponding to 
	$R_i^*=9$ in Fig. \ref{Hueco_GR}(b)). These simulations were launched at $T^*= 0.785$ and $P^*=-0.014$ (the virial pressure
	of the NVT run where the bubble configurations were generated). The bubbles grow/shrink in roughly half of the trajectories, which 
	proves that critical bubbles are obtained in the NVT simulations.}
\label{abanico}
\end{center}
\end{figure}

To test if the obtained bubbles are indeed critical, we take a number of configurations
obtained along the course of the simulation where the bubbles were stabilised and launch 
NPT simulations at the pressure given by the virial equation. 
We give an example of such test in Fig. \ref{abanico}, where we show the evolution of the bubble radius
in NPT simulations
for 30 bubble configurations with radius $\sim 9 \sigma$ obtained in the NVT simulation corresponding to 
$R_i = 9 \sigma$ in Fig. \ref{Hueco_GR}(b). 
In this and all cases the bubbles grows/shrink in roughly half of the trajectories, which is a 
clear indication that critical bubbles are equilibrated in the NVT ensemble.
Demonstrating that critical nuclei are equilibrated at constant NVT
is the most important result of this work.
\textcolor{black}{Therefore, the bubbles that minimimize F at constant
N, V and T correspond to a maximum of G at the same
N, T and at constant P (that corresponding to the
overall pressure in the system at constant V).}

\begin{table*}[]
\begin{tabular}{|c|c|c|c|c|c|c|c|c|c|c|c|c|c|}
\hline
$R_{i}$ & $R_{c}$ & $\rho_{v,dp}$ & $P_{v,dp}$ & $\rho_{l,dp}$ & $P_{l,dp}$ & $\rho_{v,cp}$ & $P_{v,cp}$ & $\Delta P$ & $N_{0}$                & $N_{T}$ & $L$                      & $V_{box}/V_{bub}$ & $\log(J)$ \\ \hline
5.7     & 6.400   & 0.0320        & 0.0207     & 0.645         & -0.0314    & 0.0345        & 0.0226     & 0.0540     & 32500                  & 32000   & 37.1026                  & 46.51             & -17.07   \\ \hline
6.8     & 5.506   & 0.0300        & 0.0197     & 0.640         & -0.0380    & 0.0340        & 0.0222     & 0.0602     & \multirow{7}{*}{32000} & 31136   & \multirow{7}{*}{36.7307} & 70.87             & -12.32   \\ \cline{1-9} \cline{11-11} \cline{13-14} 
7.6     & 7.315   & 0.0327        & 0.0211     & 0.646         & -0.0245    & 0.0355        & 0.0230     & 0.0341     &                        & 30795   &                          & 30.57             & -22.23   \\ \cline{1-9} \cline{11-11} \cline{13-14} 
8.5     & 8.466   & 0.0338        & 0.0218     & 0.648         & -0.0191    & 0.0362        & 0.0233     & 0.0342     &                        & 30342   &                          & 19.63             & -30.48   \\ \cline{1-9} \cline{11-11} \cline{13-14} 
9.4     & 9.483   & 0.0340        & 0.0219     & 0.651         & -0.0139    & 0.0369        & 0.0236     & 0.0375     &                        & 29760   &                          & 13.99             & -37.74   \\ \cline{1-9} \cline{11-11} \cline{13-14} 
10.3    & 10.48   & 0.0360        & 0.0231     & 0.652         & -0.0109    & 0.0373        & 0.0238     & 0.0374     &                        & 29034   &                          & 10.39             & -46.90   \\ \cline{1-9} \cline{11-11} \cline{13-14} 
11.2    & 11.47   & 0.0365        & 0.0234     & 0.654         & -0.0079    & 0.0376        & 0.0240     & 0.0319     &                        & 28147   &                          & 8.243             & -56.42   \\ \cline{1-9} \cline{11-11} \cline{13-14} 
12.2    & 12.44   & 0.0380        & 0.0243     & 0.655         & -0.0049    & 0.0380        & 0.0242     & 0.0290     &                        & 27082   &                          & 6.493             & -65.37   \\ \hline
12.2    & 12.13   & 0.0375        & 0.0240     & 0.653         & -0.0059    & 0.0379        & 0.0241     & 0.0300     & 136000                 & 131072  & 59.2509                  & 27.84             & -62.62   \\ \hline
\end{tabular}
\caption{Details on the simulations used to compute the bubble nucleation rate for overstretched Lennard Jones fluids. All data 
are reported in reduced units. $R_i$ is the intended
bubble radius (Eq. \ref{intended}), $R_c$ is the critical bubble radius obtained
	in NVT-Seeding by fitting the radial density profile (dp) starting at the bubble center to Eq. \ref{fit}, $\rho_{v,dp}$ and $\rho_{l,dp}$ are the
	vapor and the liquid number densities obtained from such fit, $P_{l,dp}$ ($P_{v,dp}$) is the liquid (vapor) pressure obtained from $\rho_{l,dp}$ 
	($\rho_{v,dp}$) and 
	the bulk liquid (vapor) equation of state,  
	$\rho_{v,cp}$ is the vapor density obtained by equating the chemical potential of the fluid to that of the vapor, 
	$P_{v,cp}$ is the vapor pressure obtained from $\rho_{v,cp}$ and the bulk vapor equation of state, 
	$\Delta P$ is $P_{v,cp}-P_{l,dp}$, $L$ is the
length of the simulation box edge, $N_T$ is the number of particles in the simulation box (after deletion of $N_-$ particles from a system with $N_0$ particles),  
$V_{box}/V_{bub}$ is the volume ratio between the box and the bubble, 
	and $J$ is the nucleation rate.}
	\label{tabla}
\end{table*}

\subsection{Calculation of $\Delta P$}
\begin{figure}[h]
\begin{center}
	\includegraphics[width=0.85\linewidth]{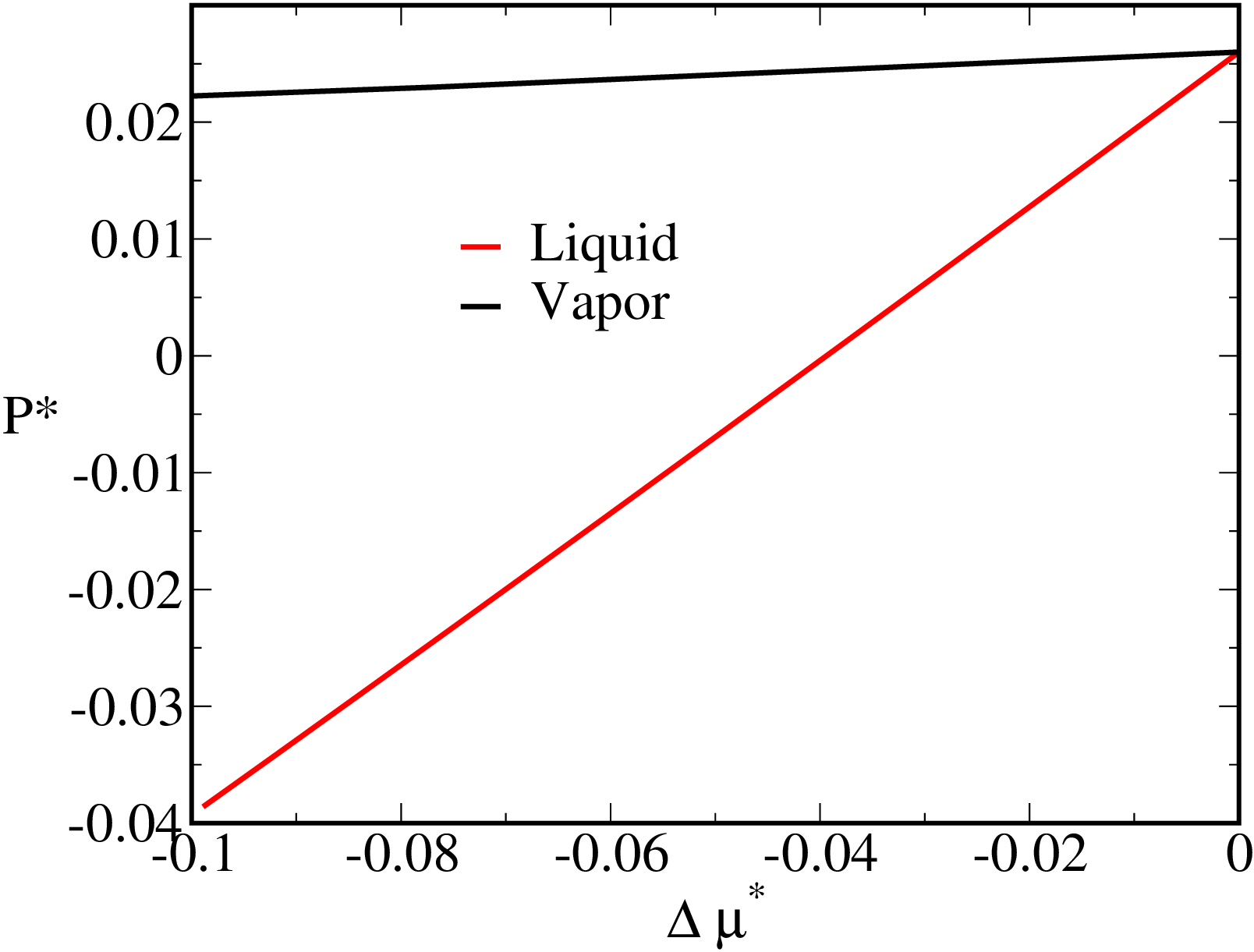}(a)
	\includegraphics[width=0.85\linewidth]{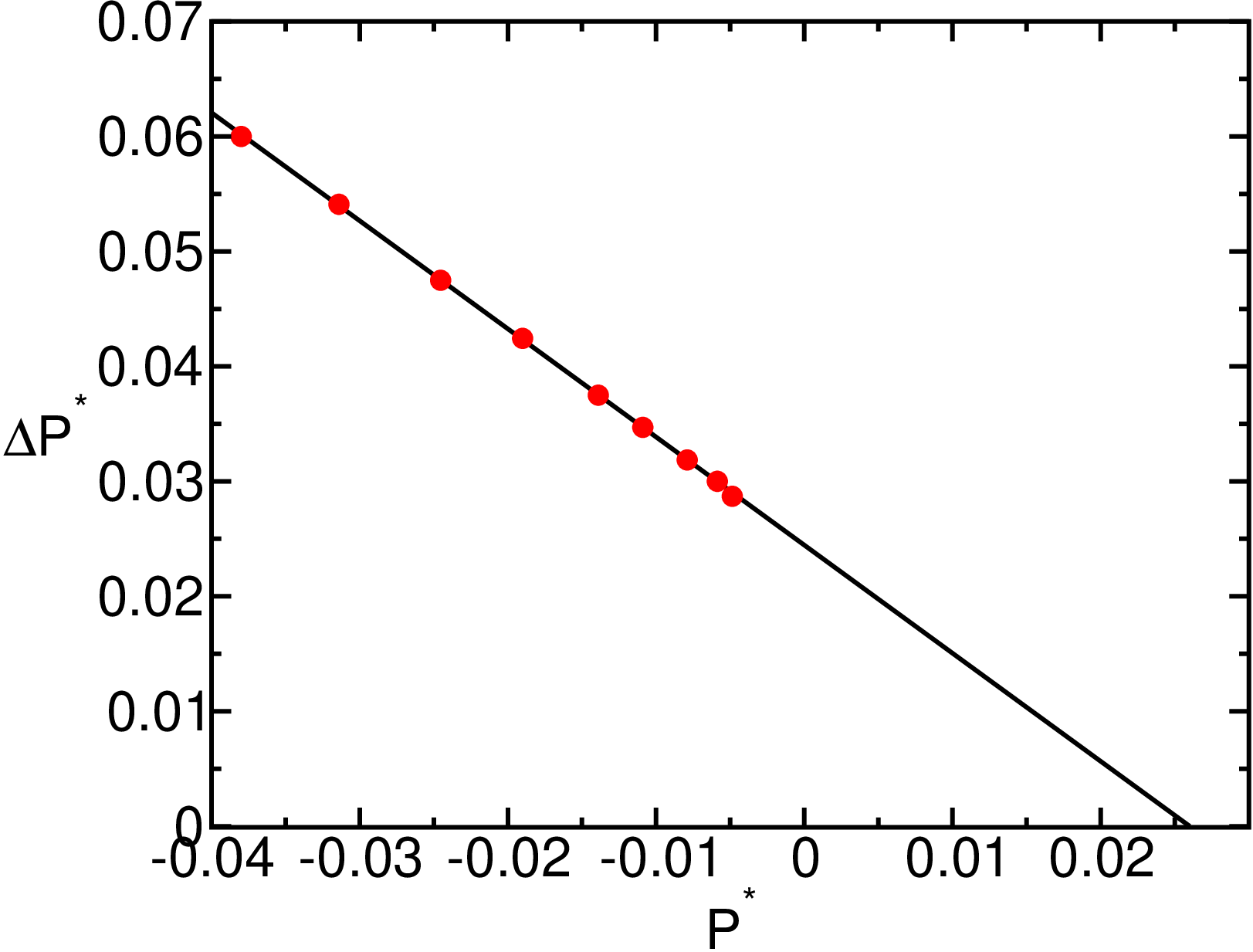}(b)
	\caption{(a) Pressure versus the chemical potential  
		difference with respect to coexistence as obtained from Eq. \ref{TI} for the liquid and the vapor. (b) $\Delta P$ versus
		the liquid pressure. Red dots are the values of $\Delta P$ for the bubbles studied in this work.} 
\label{DeltaP_GR}
\end{center}
\end{figure}

$\Delta P$ is the pressure difference between the vapor inside the critical bubble and that of the 
surrounding liquid. 
We obtain the liquid pressure through $\rho_{l,dp}$, the liquid density from the density profiles, and the bulk liquid equation of state
that relates pressure with density (red dots in Fig. \ref{vm}). 
Such pressure is reported in Table \ref{tabla} as $P_{l,dp}$. 
To find the vapor pressure inside the critical bubble, 
we follow our previous approach \cite{seedingNpT} and 
use the fact that the chemical potential of the liquid and the vapor phases are equal both at coexistence and at 
nucleation conditions \cite{tolman1949effect}. 
The chemical potential difference with respect to coexistence at any 
pressure can be found by integrating the inverse density from the coexistence pressure, ($P^*_{coex}=0.026$ at $T^*=0.785$):
\begin{equation}
	\Delta \mu (P) = \mu (P) - \mu_{coex} = \int_{P_{coex}}^{P} \frac{1}{\rho}dP 
\label{TI}
\end{equation}
In Fig. \ref{DeltaP_GR} we show in black and red the pressure versus the chemical potential
difference with coexistence for the vapor and the liquid
respectively. 
For a given liquid pressure ($P_{l,dp}$) we obtain the vapor pressure by reading in Fig. \ref{DeltaP_GR}(a) 
the vapor pressure for $\Delta \mu (P_{l,dp})$.
We refer to this vapor pressure, obtained by chemical potential (cp) equality, as $P_{v,cp}$.
Finally, $\Delta P$ is $P_{v,cp}-P_{l,dp}$. 
The values of $\Delta P$ thus obtained for the bubbles studied in this work are reported in Table \ref{tabla} and plotted versus
$P_{l,dp}$ in Fig. \ref{DeltaP_GR} (b) with red dots.

We could have obtained the vapor pressure via $\rho_{v,dp}$ 
(the density inside the bubble) and the vapor equation of state. The resulting vapor pressure, 
$P_{v,dp}$, is reported in table \ref{tabla}.
We do not use $P_{v,dp}$ to compute $\Delta P$
because such vapor pressure value does not 
guarantee that the vapor chemical potential is equal to that of the surrounding liquid. In table \ref{tabla} it can be seen that 
$P_{v,dp}$ is similar to $P_{v,cp}$, although systematically smaller.
\textcolor{black}{The difference arises from the discrepancy between the
actual density near the center of the bubble and that of the
hypothetical system. Consistently,  
the discrepancy increases as stretching (or the
superheating) is increased and the bubbles become smaller.}
Therefore, computing the bubble pressure through 
$\rho_{v,dp}$ gives reasonable but not accurate or rigorous values.

In order to obtain a function that gives $\Delta P$ at any liquid pressure --which will be needed to fit $J$ along pressure-- we do as follows:
we fit the $P(\Delta \mu)$ data shown in Fig. \ref{DeltaP_GR}(a) to a linear fit for the liquid and for the vapor. 
The difference between both fits gives $\Delta P$ as a function 
of $\Delta \mu$. Then, we substitute $\Delta \mu$ in the resulting expression by 
$\Delta \mu(P_{liquid})$. This gives $\Delta P (P_{liquid})$, that we represent in Fig. \ref{DeltaP_GR}(b) with a black
line. 

\subsection{Nucleation rate}
With Eq. \ref{rateeq} and the values reported in Table \ref{tabla} 
for $\Delta P$, $\rho_{l,dp}$ and $R_c$ we obtain the nucleation rate, 
reported also in Table \ref{tabla}.
The NVT-Seeding results for the nucleation rate are shown with 
pink solid circles in Fig. \ref{Tasa_GR}.
These are compared with Seeding results in the NPT ensemble (brown dots), obtained as described in 
Ref. \cite{seedingNpT}. The agreement between Seeding results in both ensembles is excellent. 
In Fig. \ref{Tasa_GR} we also include with orange diamonds data obtained for the bubble nucleation rate in conditions
of high overstretching where cavitation is spontaneous in a simulation starting from a bulk liquid configuration. 
In such conditions the rate can be estimated as $J=1/(Vt)$ \cite{filion:244115}, where $t$ is the average nucleation time 
in a handful of independent trajectories and $V$ is the system's volume. The results for the nucleation rate obtained by brute force 
molecular dynamics are reported in table \ref{brute}. 
Although Seeding cannot be directly used in conditions where cavitation is spontaneous
given that the critical bubbles are too small, the trend of Seeding data is consistent with the spontaneous cavitation data.  

\begin{table}
\begin{tabular}{|l|l|l|l|}
\hline
$P^{*}$      & -0.090 & -0.082 & -0.075 \\ \hline
$log(J^{*})$ & -6.33  & -6.59  & -6.95  \\ \hline
\end{tabular}
\caption{Nucleation rate in conditions of high overstretching, where the bubbles spontaneously nucleate in a brute force molecular dynamics simulation. All simulations
have been run with a system of 32000 particles.}
\label{brute}
\end{table}

In a recent publication, we showed that when the bubble radius is measured with the definition adopted in 
this paper, NPT-Seeding gives bubble nucleation rates consistent with those computed by means of independent 
rare event simulation techniques \cite{seedingNpT}. Here we demonstrate that NVT and NPT-Seeding give the same results. 
The advantage of NVT over NPT-Seeding is that the critical bubble is naturally equilibrated along the course of
the simulation. 
In the NPT ensemble, however, the bubble either grows, if the pressure is lower than that for which the inserted bubble is critical,  or shrinks 
in the contrary case. 
Thus, by performing simulations at different pressures, 
the pressure that makes the bubble critical is enclosed within a certain range. 
This procedure is much more cumbersome than just letting the critical bubble equilibrate by itself as we do in the NVT ensemble. 
Moreover, NPT-Seeding entails an error in the pressure that makes the bubble critical.
By contrast, in the NVT ensemble the pressure  is obtained from the density 
of the fluid surrounding the bubble, that can be accurately averaged 
along the course of the long NVT simulation in which the critical bubble is stable. 
Therefore, we recommend the NVT ensemble to compute bubble nucleation rates along isotherms via Seeding.

The only caveat for the use of Seeding in the NVT ensemble is the appearance of finite size effects
when the bubble is big as compared to the simulation box. 
In practice, one should check that the fluid reaches a plateau density at
$L/2$ from the bubble centre. In Fig. \ref{bubblecenter}
we show that the two largest bubbles generated in a box with $L^*=36.73$ do not meet this requirement, so we anticipate that
these systems might be affected by finite size effects.
The rate corresponding to these systems is shown with empty circles in Fig.\ref{Tasa_GR}.
Finite size effects are not strong given that
neither data clearly deviates from the general trend.
We find that NVT-Seeding is a very promising strategy to enhance the efficiency and accuracy of Seeding. 
In our study finite size effects started to appear when the volume ratio between the simulation box and the critical bubble was 
smaller than $\sim 8$ (see Table \ref{tabla}). Further work is required to establish if the appearance of finite size effects in NVT-Seeding 
below this volume ratio is more general. 

\subsection{Nucleation rate fit and surface tension}

\begin{figure}[h]
\begin{center}
\includegraphics[width=0.85\linewidth]{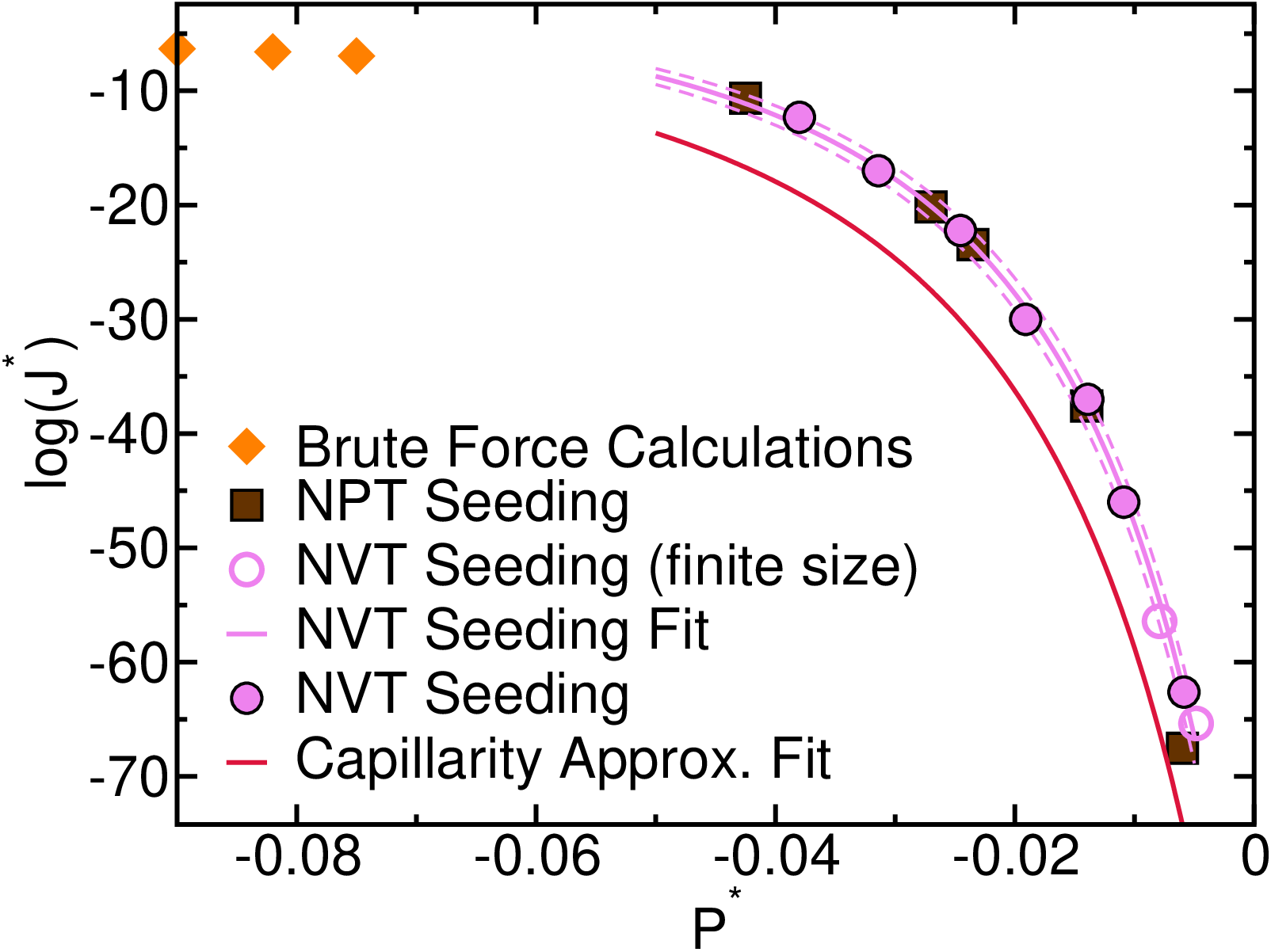}
\caption{Nucleation rate as a function of the liquid pressure at constant temperature $T^*=0.785$. 
	Pink and brown symbols correspond to Seeding data obtained in the NVT and NPT ensemble respectively.
	Empty symbols are data subject to finite size effects. 
	Brute force calculations are shown as orange diamonds.
	Pink curve is a CNT fit (Eq. \ref{rateeq}) to the NVT seeding data. The dashed pink curves indicate the upper and lower error limits in 
the solid one by considering an error of 0.1 $\sigma$ in $R_c$ and a relative error of 0.5 per cent in $\Delta P$. 
	Red curve corresponds to a $J$ estimate based on the capillarity approximation (Eq. \ref{rateeq} assuming that $\gamma$ is equal to the 
	coexistence $\gamma$ for any pressure). 
	}
\label{Tasa_GR}
\end{center}
\end{figure}

\begin{figure}[h]
\begin{center}
\includegraphics[width=0.85\linewidth]{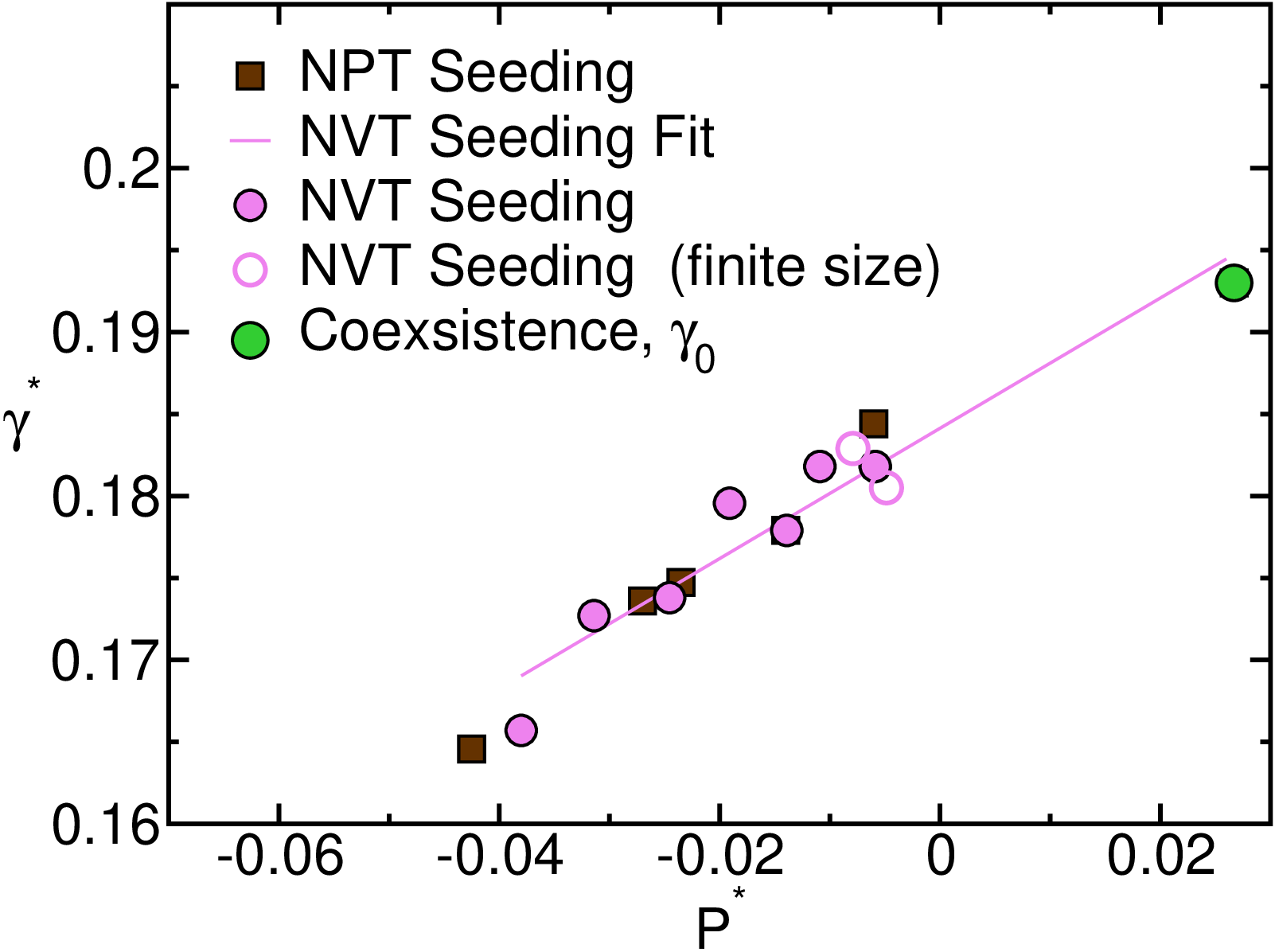}(a)
\includegraphics[width=0.85\linewidth]{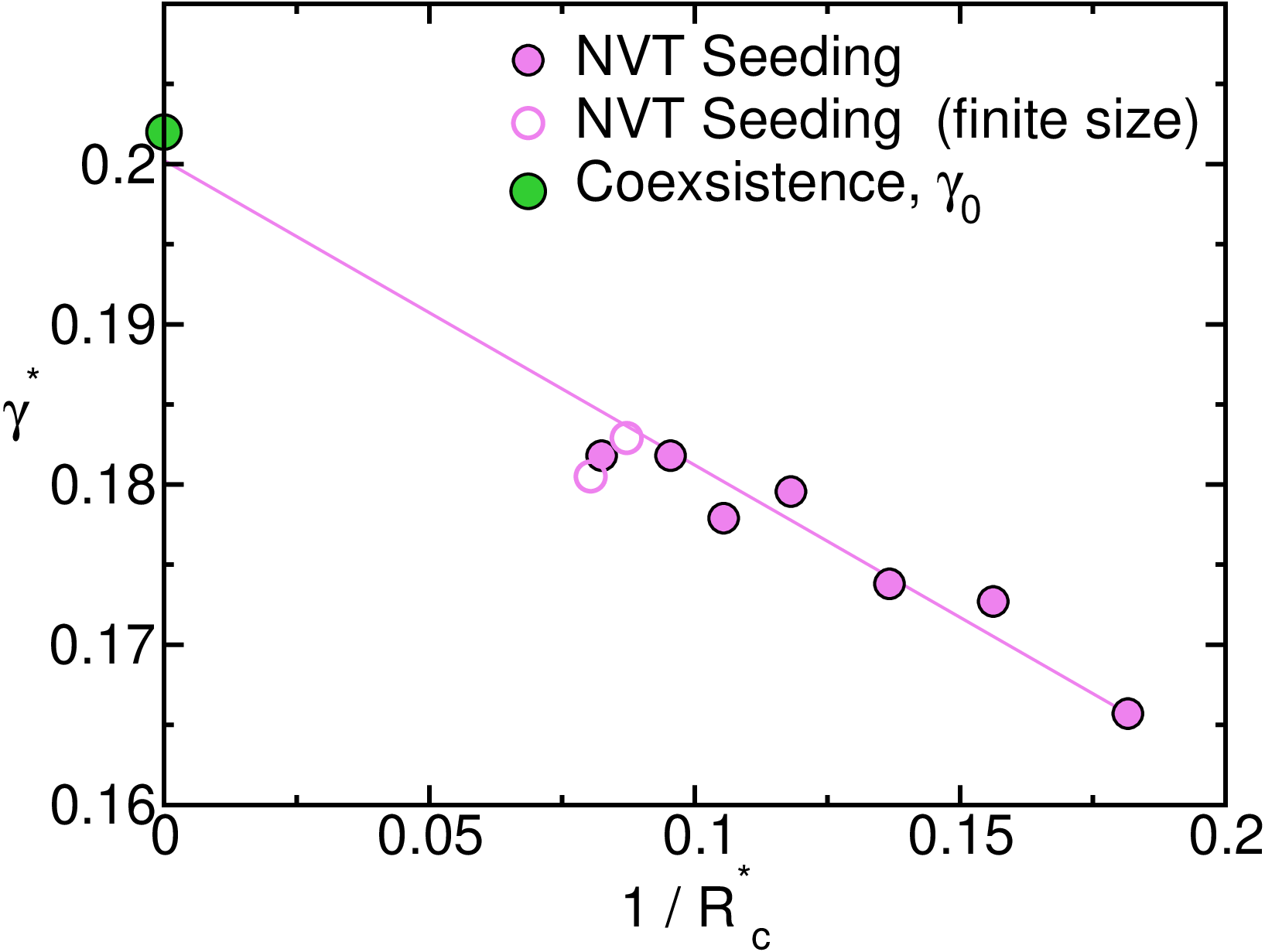}(b)
\caption{(a)Interfacial free energy as a function of the liquid pressure. Pink (brown) circles (squares) correspond to NVT (NPT) Seeding simulations. 
	The coexistence $\gamma$, shown with a green circle, was obtained in Ref. \cite{seedingNpT}. 
	The pink line is a linear fit to the coexistence value and the NVT-Seeding data. (b) $\gamma$ versus the inverse critical radius.}
\label{Gamma_GR}
\end{center}
\end{figure}

The pink curve in Fig. \ref{Tasa_GR} is a CNT inspired fit to the NVT-Seeding data. 
To get such curve we
need, according to Eq. \ref{rateeq}, $\Delta P$, $\rho_l$ and $R_c$ as a function of pressure. 
$\rho_l$ and $\Delta P$ versus pressure have already been presented in Figs. \ref{vm} and \ref{DeltaP_GR} respectively. 
We have already evaluated $R_c$ for different pressures from the seeding simulations (see table \ref{tabla}). However, 
the dependence of $R_c$ with the liquid pressure is far from linear. Rather than directly fitting $R_c$ vs the liquid pressure, 
it is more convenient to get
the surface tension dependence with pressure first through the Laplace equation $\gamma=\Delta P R_c/2$.  
The $\gamma$ values obtained from the $R_c$ and $\Delta P$ Seeding data combined with the Laplace equation are shown in Fig. \ref{Gamma_GR}(a).
As expected from the good agreement for the nucleation rate, NPT and NVT-Seeding give the same $\gamma$.
We observe that $\gamma$ decreases 
as the bubbles become smaller, 
in contrast with Ref. \cite{matsumoto2008nano} where  $\gamma$ was found to be roughly constant
in NVT simulations of bubbles.
The NVT-Seeding $\gamma$ data can be linearly fitted alongside the coexistence $\gamma$, $\gamma_0$ (green dot in the figure). 
We take $\gamma_0$ from our previous work 
\cite{seedingNpT}. 
With $\gamma(P)$ and the Laplace equation we obtain $R_c$ at any liquid pressure. 
With that, we have all the ingredients required to fit $J$
(pink curve in Fig. \ref{Tasa_GR}).

If $\gamma$ is assumed to be equal to the coexistence value for any pressure (capillarity approximation)
one gets the red curve in Fig. \ref{Tasa_GR}. As expected, the red curve is below the pink one given
that the coexistence $\gamma$ is larger than those obtained by Seeding for overstretched fluids. 
Therefore, a theoretical description based on CNT and the capillarity approximation fails in predicting bubble nucleation rates. 

Finally, we represent the $\gamma$ data shown in Fig. \ref{Gamma_GR}(a) 
as a function of $1/R_c$ in Fig. \ref{Gamma_GR}(b).
The straight line is a fit to the following expression proposed by Tolman \cite{tolman1949effect}:
\begin{equation}
\gamma=\gamma_0 \left(1-2\delta_T/R_c\right),
\label{tolmanfit}
\end{equation}
where $\delta_T$ is the Tolman length, which sign tells if $\gamma$
increases or decreases with curvature at constant temperature and its magnitude indicates how strong is such variation \cite{kashchievbook,blokhuis2006thermodynamic,wilhelmsen2015tolman,sampayo}. 
In this case we obtain a positive $\delta_T$ (hence $\gamma$ decreases with curvature) of about the 
particle radius: $\delta_T^*=0.47$. 

\textcolor{black}{It is fair to point out here that the theoretical framework employed in this paper (the Laplace equation and Eqs. \ref{rateeq} and \ref{tolmanfit}) assumes that the critical bubble radius is that corresponding to the surface of tension. 
In principle, such radius does not necessarily coincide with that obtained in our simulations as explained in section \ref{cbr_sec}. However, in a recent study we found that such radius definition plus CNT provides free energy barrier heights consistent with those obtained from independent Umbrella Sampling calculations \cite{seedingvienes}. This suggests that our radius definition provides a good estimate of the radius of tension. This idea is supported in this paper by the consistency between seeding and brute force data for the nucleation rate (see Fig. \ref{Tasa_GR}).}

\section{Summary and conclusions}

In this paper we study the nucleation of bubbles in overstretched Lennard Jones fluids at constant temperature.  
We first ``seed'' the fluid with a cavity 
by randomly removing particles and switching on a repulsive step-like potential. 
We then switch off the respulsive potential and let the system evolve at constant 
volume and temperature (NVT ensemble). The cavity becomes a vapor bubble in equilibrium with the surrounding liquid. 
We demonstrate that such bubble corresponds to the critical 
one at the thermodynamic conditions
of the surrounding liquid. We measure the bubble radius by computing radial 
density profiles from its center and the vapor pressure by means of thermodynamic
integration considering equal chemical potential between the bubble and the liquid. 
With the bubble radius and the pressure difference between both phases, we estimate the 
bubble nucleation rate using Classical Nucleation Theory to compute the free energy barrier height alongside
an expression provided by Blander and Kats to estimate the kinetic pre-factor (Eq. \ref{rateeq}).
We name this approach to obtain the nucleation rate NVT-Seeding. 
To fit the computed nucleation rates we needed to consider a pressure dependent surface tension. Therefore, the capillarity approximation is not
valid to make theoretical predictions of bubble nucleation. 

The nucleation rate obtained by NVT-Seeding, here used for the first time, is fully consistent with that obtained by Seeding at constant pressure,  
NPT-Seeding, which is the
approach adopted so far to study either bubble \cite{seedingNpT} or  
crystal nucleation \cite{knottJACS2012,jacs2013,seedingvienes,petersJACS2015,espinosaJCP2014,espinosaJCP2016,espinosaPRL2016,espinosaJPCL2017,guiomar,sossoJCP2016} via Seeding. 
NVT-Seeding enables
a more accurate computation of the critical bubble parameters than its NPT counterpart because the bubble remains stable
along the course of the simulation. Moreover, given that the bubble equilibrates by itself, 
it is less cumbersome to apply. NVT-Seeding is quite handy to obtain the nucleation rate along pressure for a given temperature. 
However, to obtain the rate along an isobar NPT-Seeding is more suitable since in NVT-Seeding the pressure cannot be easily controlled because 
it is coupled to the bubble radius, which changes during the equilibration 
of the generated cavity. There are also size limitations in NVT-Seeding: 
too small bubbles will dissolve and too large ones will suffer from finite 
size effects due to the lack of surrounding liquid. 
More work is required to establish the applicability limits of the NVT-Seeding technique and to test if it is also valid to study nucleation 
in other sorts of phase transitions such as freezing or condensation.

\section{Acknowledgments}
This work was funded by grant FIS2016/78117-P of the MEC.
C. Valeriani thanks financial support from FIS2016-78847-P of the MEC.
The authors acknowledge the computer resources and technical assistance provided by the RES. 
P. R. thanks a doctoral grant from UCM.



\end{document}